\documentclass[a4paper,fleqn]{cas-sc}
\usepackage[numbers]{natbib}
\usepackage{float}
\usepackage{graphicx}
\graphicspath{figures/}
\usepackage[justification=centering]{caption}
\usepackage{subfig}
\usepackage{hyperref}
\usepackage{multicol}
\usepackage{multirow}
\usepackage{romannum}
\usepackage{array}
\usepackage{amsmath}

\begin{document}

\let\WriteBookmarks\relax
\def\floatpagepagefraction{1}
\def\textpagefraction{.001}

\shorttitle{Radiopurity studies in Aut rock}    

\shortauthors{Swati Thakur et~al.}  

\title [mode = title]{Radiopurity studies of a rock sample from the Aut region} 

\affiliation[1]{organization={Department of Physics, Indian Institute of Technology Ropar, Rupnagar, Punjab - 140001, India}}
\affiliation[2]{organization={TIF-AWaDH, Indian Institute of Technology Ropar, Rupnagar, Punjab - 140001, India}}
\affiliation[3]{organization={Department of Nuclear and Atomic Physics, Tata Institute of Fundamental Research, Mumbai - 400005, India}}
\affiliation[4]{organization={India-based Neutrino Observatory, Tata Institute of Fundamental Research, Mumbai - 400005, India}}
\affiliation[5]{organization={Homi Bhabha National Institute, Anushaktinagar, Mumbai - 400094, India}}
\affiliation[6]{organization={Pelletron Linac Facility, Tata Institute of Fundamental Research, Mumbai - 400005, India}}
\affiliation[7]{organization={Department of Physics, Himachal Pradesh University, Shimla - 171005, India}}

\author[1]{Swati Thakur}
\author[2]{A. Mazumdar}
\author[3]{Nishant Jangid}
\author[4,5]{V. Vatsa}
\author[3]{M.S. Pose}
\author[3]{S. Mallikarjunachary}
\author[6]{S. Pal}
\author[3]{V. Nanal} [orcid=0000-0002-2122-5191]
\cortext[1]{V. Nanal}
\ead{nanal@tifr.res.in}
\author[1]{R.G. Pillay}
\author[1]{P.K. Raina}
\author[1]{Pushpendra P. Singh}
\author[7]{S.K. Dhiman}

\begin{abstract}
Efforts are underway to set up an underground laboratory in India for rare event studies like double beta decay, dark matter, etc. For such experiments, mitigation of radiation background is of paramount importance and understanding ambient background at the site, originating from the rock, is one of the crucial factors. With this motivation, the radiopurity studies of a rock sample from the potential laboratory site in the Aut tunnel of Himachal Pradesh (India) have been carried out using the TIFR low background experimental setup (TiLES). The concentration of $^{40}$K in Aut rock is observed to be lower by a factor of $\sim$\,1000 as compared to the samples from BWH (Bodi West Hill), Tamil Nadu (India), current designated site for India-based Neutrino Observatory. The natural radioactive trace impurity $^{232}$Th is lower in the Aut rock, while $^{238}$U is somewhat higher than the BWH rock. 
Overall, the ambient gamma ray background at Aut is expected to be lower than the BWH, while ambient neutron background is expected to be similar. Further, to assess the neutron-induced long lived activity, fast neutron activation studies have been carried out on the both Aut and BWH rock samples at the Pelletron Linac Facility, Mumbai. 
\end{abstract}

\begin{keywords}
 Low background measurements\sep Gamma ray spectroscopy\sep Neutron activation studies 
\end{keywords}

\maketitle

\section{Introduction}\label{Introduction}
Recently, experimental investigations of physics beyond the standard model, such as neutrino oscillations, neutrinoless double beta decay (NDBD), dark matter search, etc., have attracted much attention worldwide~\cite{Yasaman2018,Roszkowski2018,Dolinski2019}. These studies for rare events demand stringent background levels. The ultimate background levels ($N_{bkg}$) achieved  in the region of interest are 4.0\,$\times$\,10$^{-4}$\,keV$^{-1}$kg$^{-1}$y$^{-1}$ for KamLAND-Zen NDBD experiment~\cite{ostrovskiy2016} and 8.5\,$\times$\,10$^{-2}$\,keVee$^{-1}$kg$^{-1}$y$^{-1}$ for  XENON1T dark matter experiment~\cite{Aprile2018}. It is important to note that  minimizing the ambient background is a crucial aspect for rare event studies.
Some of the major sources of the background radiation are cosmic rays, long-lived primordial radionuclides ($T_{1/2}$ $\sim$\,$10^{8}$ - $10^{10}$\,y), cosmogenic radionuclides ($T_{1/2}$ $\sim$\,days - years) and neutron induced activity produced in and around the detector~\cite{Dolinski2019}. In order to suppress the cosmic muon background (by about 5 - 6 orders of magnitude), these experiments are located in underground laboratories, typically with a rock overburden of more than 500\,m. In an underground laboratory, the ambient gamma and neutron background at the site and the background arising from penetrating (high energy) cosmic muons, can be the limiting factors for the sensitivity of the experiment.

The gamma ray background originates from natural radioactivity due to trace elements like $^{40}$K, $^{232}$Th, $^{235,238}$U in the rock, and the neutron induced activities of the constituents of the rock. The concentrations of these trace elements, and subsequently the associated gamma/neutron backgrounds, are known to exhibit geographical variation, depending on local geological conditions. Amongst the natural gamma ray background, high energy gamma rays ($E$\,$\ge$\,2\,MeV) produced in the natural radioactivity chains of $^{238}$U and $^{232}$Th e.g. 2448\,keV ($^{214}$Bi) and 2615\,keV ($^{208}$Tl), respectively, are of significant concern. The neutron background is usually categorized in two parts -low energy neutrons ($E_n$\,<\,20\,MeV), originating from the spontaneous fission of uranium and ($\alpha$, n) reactions in the rock, and very high energy neutrons ($E_n$\,>\,1\,GeV) produced by muon-induced interactions in the rocks or materials surrounding the detector. The flux of high energy neutrons is expected to be smaller by a factor of 10$^2$ - 10$^3$ as compared to the low energy neutrons~\cite{Mei2006}. The inelastic scattering of neutrons (n, n$^{\prime}\gamma$) and neutron-capture (n, $\gamma$) with the detector and/or surrounding materials can also lead to the production of high energy gamma rays. The neutron-activated reaction products can have wide-ranging half-lives ($\sim$\,min to $\sim$\,years) and the contribution of long-lived activities is difficult to mitigate.

In India, driven by the interest in rare decay studies, a proposal for an underground laboratory has been initiated. A laboratory with about 1\,km rock overburden is proposed in  Bodi West Hills (BWH) of the Theni district in Tamil Nadu (Lat. North 9$^{\circ}$57'47.65'' and Long. East 77$^{\circ}$16'22.55'')~\cite{mondal2012india}. Also, a small laboratory (approximately 5\,m\,$\times$\,5\,m\,$\times$\,2.2\,m) has been set up at UCIL, Jaduguda, Jharkhand with a rock cover of 555\,m inside an Uranium mine~\cite{banik2021simulation}.
Another potential site with a reasonable rock overburden ($\sim$\,500\,m) has been identified in the existing Aut tunnel, Himachal Pradesh (Lat.~North 31.725$^{\circ}$ and Long. East 77.206$^{\circ}$).
The Aut rock is Dolomite type, while the BWH rock is Charnockite. To assess the feasibility of this site, the radiation background studies are important and hence the radiopurity studies of the Aut rock samples have been carried out. The concentrations of trace level natural radioactive elements are determined and are compared with the BWH rock sample data. Additionally, the neutron activation measurements have been performed for both Aut and BWH rock samples, with an emphasis to study long-lived activities.

The paper is organized as follows: Experimental details for the radiopurity and neutron activation measurements are described in section~\ref{Experimental details}. The analysis of the radiopurity measurements and the estimation of trace level radioimpurities are discussed in section~\ref{Radiopurity measurements}. The results of the neutron induced activities are presented in section~\ref{Neutron activation measurements}. Finally, the results are summarised in section~\ref{Summary and discussions}. 

\section{Experimental details}\label{Experimental details}
The rock samples used in the present study were collected from the Aut tunnel (from the location of the proposed laboratory site) and the BWH site. The latter sample was bored from a depth of $\sim$\,30\,m. The densities of the Aut rock sample and the BWH rock were measured to be $\sim$\,2.93\,g/cm$^{3}$ and $\sim$\,2.89\,g/cm$^{3}$, respectively. The gamma ray spectra were measured using two low background counting setups at TIFR (Mumbai) at sea level - TiLES (TIFR Low background Experimental Setup)~\cite{Dokania2014} and a coincidence setup of two low background HPGe detectors (D1-D2)~\cite{thakur2021icwip}. Neutron activation studies were performed at the fast neutron irradiation setup at the Pelletron Linac Facility (PLF), Mumbai. 

\subsection {Low background counting setups}
The TiLES consists of a $\sim$\,70\,\% relative efficiency high-purity Germanium (HPGe) detector (ORTEC, Model no GEM75-95-LB-C-HJ) in a passive shielding of 10\,cm thick low activity Pb ($<$\,0.3\,Bq/kg of $^{210}$Pb) and 5\,cm oxygen-free high thermal conductivity Cu. 
The setup has a provision to further reduce the ambient background by employing a cosmic muon veto and continuous dry nitrogen flushing. However, the cosmic muon veto and dry nitrogen flushing were not used for the present measurements as the samples had sufficiently high activity compared to the ambient background as well as to facilitate frequent sample changes. The D1-D2 coincidence setup consists of two identical low background cryocooled HPGe detectors (ORTEC, Model no GEM30P4-83-RB). Both detectors have $\sim$\,33\,\% relative efficiency and have a carbon fiber housing with a 0.9\,mm thick front window. The D1-D2 detectors are mounted in a close geometry, facing each other at a distance of about 2.5\,cm, surrounded by two layers of passive shielding with 5\,cm thick  
low activity lead ($<$\,0.3\,Bq/kg of $^{210}$Pb) inside and 5\,cm thick ($<$\,21\,Bq/kg of $^{210}$Pb) outside. As both D1 and D2 are expected to be identical, the optimized geometry of D1~\cite{gupta2018} was adopted for simulation of the photopeak efficiency of both the detectors (D1 and D2).

\begin{figure}[h]
\centering
\includegraphics[width=12.0cm,height=10.0cm]{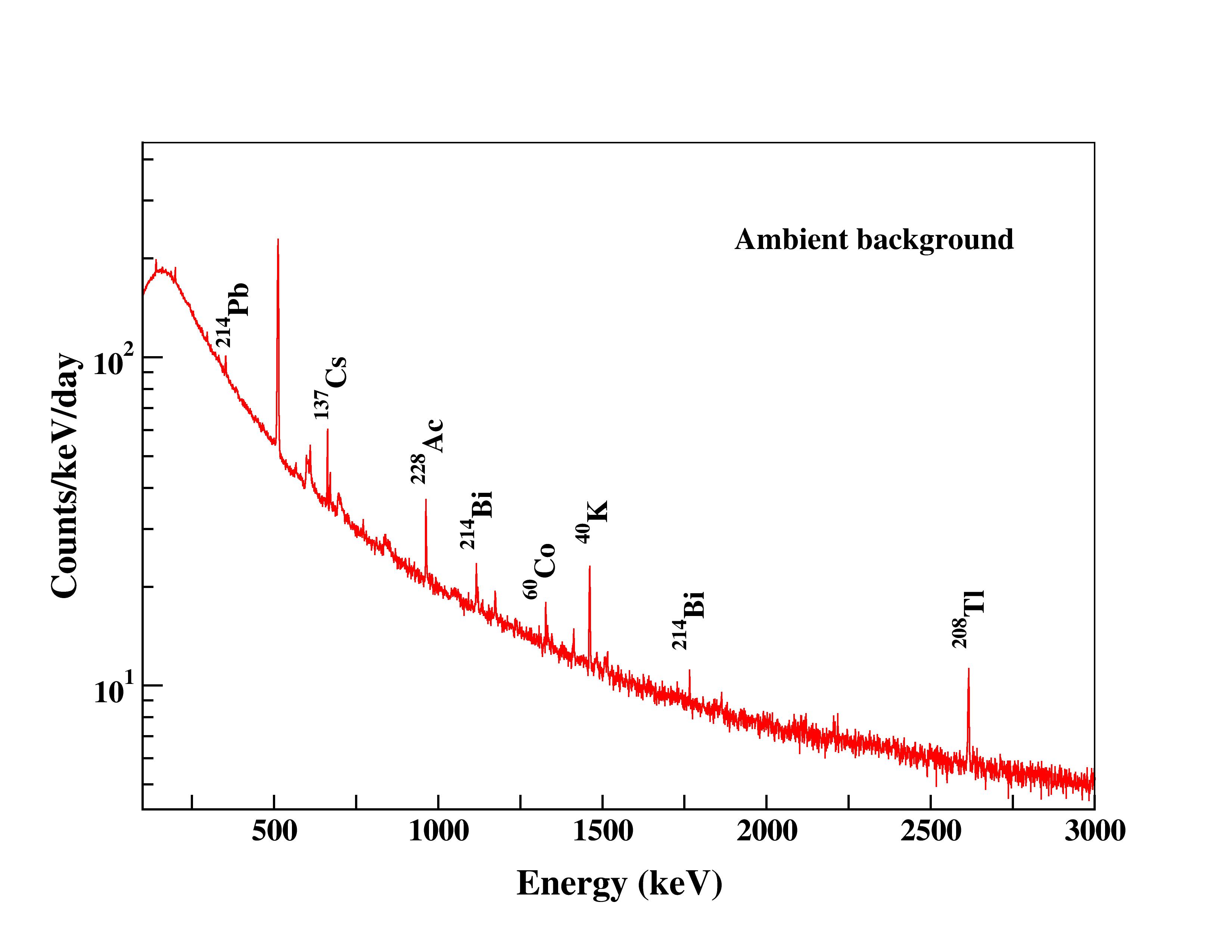}
\caption{A typical gamma ray spectrum of the ambient background in the TiLES, where the prominent gamma rays are labelled (counting time $t$\,=\,69\,d).}\label{fig1}
\end{figure}

Data were recorded using a 14-bit, 100\,MHz CAEN 6724 series commercial digitizers using the trapezoidal filter for pulse height determination. The digital parameters were optimized to achieve the best energy resolution. For both setups, the optimum trapezoidal filter settings, namely, input signal decay time ($T_{decay}$), trapezoidal rise time ($T_{rise}$) and trapezoidal flat top time ($T_{flattop}$) were set at 50\,$\mu$s, 5.5\,$\mu$s and 1.3\,$\mu$s, respectively. The typical resolution at 1460\,keV was $\sim$\,2.6\,keV in TiLES and $\sim$\,3.5\,keV for D2.
The data were analyzed offline using ROOT~\cite{root} and LAMPS~\cite{lamps} software. The energy calibration was obtained using standard gamma ray sources prior to the measurement and were monitored with known background lines over an energy range of 120 - 2615\,keV. The observed drifts were less than 1\,keV over an extended period of one year. The dead time was monitored with a standard 10\,Hz pulser and was found to be negligible ($<$\,0.1\,\%).  
Figure~\ref{fig1} shows the background measured in TiLES (with the perspex sample mounting plate) for a total duration of 69\,d, acquired at different times over a period of about 7 months.

For radiopurity measurements both the Aut and BWH rock samples were mounted on the sample plate at $d$\,$\sim$\,10\,mm from the face of the detector in TiLES. The details of the samples, namely, mass and counting time are given in Table~\ref{tbl1}. The photopeak efficiencies ($\epsilon$) for various gamma rays in each rock sample were obtained using GEANT4 based simulation program~\cite{Agostinelli2003} for respective counting configurations. Since the rocks were of irregular shape, the geometry of the sample was simulated by approximating the closest regular shapes corresponding to the sample volume. The gamma rays were assumed to originate from trace impurities uniformly distributed within the sample. For each  energy, $10^6$ (1\,M) gamma rays were generated within the sample isotropically and consequently statistical errors in simulated photopeak efficiencies were negligible ($<$\,1\,\%). 
Further, the uncertainties in the detection efficiency due to variation in rock dimensions about mean values were estimated. Since the largest linear dimension of both samples were smaller than the detector crystal size, only the variation in thickness is expected to have a pronounced effect on the efficiency. 
The thickness variation over the sample size  was found to be 8.1 - 8.6\,mm for AUT0 and 9 - 12\,mm for BWH0. In simulation, different shapes were generated corresponding to different thicknesses, by appropriately modifying the cross-sectional area to keep the volume constant. The $\epsilon_\gamma$ is taken as average value of $\epsilon$ for different shapes, while the  error in the efficiency  is calculated as ($\epsilon_{max}$-$\epsilon_{min}$)/2. The overall observed spread in efficiency due to shape variation  was $<\,\pm\,5\,\%$, for the energy range of interest ($E_\gamma=186 - 2615$\,keV). Another important aspect that needs to be taken into consideration, especially for close counting geometry, is the coincident summing of multiple gamma rays in the decay cascade. While detailed simulations are essential for multi-stage decay cascades, a simple estimation can be done for a two gamma cascade $2\xrightarrow{\gamma_2}1 \xrightarrow{\gamma_1} 0$.
If $\epsilon_i^{tot}$ is the total detection probability of $\gamma_{i}$ (i.e. Compton and photopeak), then the loss due to the coincident summing ($I_{1}^{sum}$, $I_{2}^{sum}$) to the intensity of  $\gamma_1$  and $\gamma_2$ ($I_{\gamma 1}$ and $I_{\gamma 2}$, respectively) can be  estimated as   

\begin {equation}\label{eq:1}
\begin{aligned}
I_{1}^{sum} =  \epsilon_2^{tot} \cdot p_1  \cdot p_2\cdot f_2 \\
I_{2}^{sum} =  \epsilon_1^{tot} \cdot p_1  \cdot p_2\cdot f_2
\end{aligned}
\end {equation}
where $p_1$ and $p_2$ are the gamma decay probabilities of levels 1 and 2, respectively and $f_2$ is the feeding fraction for the level 2. The effective net intensity for $\gamma_i$ can be written as

\begin {equation}\label{eq:2}
I_{i}^{net} = I_{\gamma i}-I_{i}^{sum}
\end {equation}

Here, both gamma rays are assumed to be emitted isotropically, neglecting angular correlations. 
It is evident that the coincident summing probability of 3 or more gamma rays is insignificant. However, the  complex decay cascades with multiple parallel decay branches are not considered for estimating elemental concentration.  
It should be mentioned that the systematic errors in the simulated efficiencies of the optimized models for TiLES and D1/D2 are 5\,\% and 8\,\%, respectively. 
Total uncertainties quoted include all contributions, namely, statistical, systematic (due to detector modelling) and coincident summing.

\begin{table}[width=1.0\linewidth,cols=4,pos=h]
\centering
\caption{ Details of the rock samples for radiopurity study in TiLES}\label{tbl1}
\begin{tabular*}{\tblwidth}{@{}CCCC@{}}
\toprule
Sample      & Type          & Mass                 & Counting time \\
            &               & (g)                  & (d)        \\
\midrule
BWH0        & Charnockite   & 32.2                 & 24.9    \\
AUT0        & Dolomite      & 27.5                 & 23.4    \\
\bottomrule
\end{tabular*}
\end{table}

\subsection{Neutron activation}
For neutron activation measurements, fast neutrons having a broad energy range upto $\sim$\,20\,MeV were produced at the irradiation setup of the Pelletron Linac Facility, Mumbai, using $^{9}$Be(p,n)$^{9}$B reaction ($Q$ = -1.85\,MeV) with $\sim$\,5\,mm thick $^{9}$Be target~\cite{Sabyasachi2014}. The maximum proton beam energy (22\,MeV) was chosen to cover the energy range of neutron spectra originating from fission and ($\alpha$, n) reactions in the rocks~\cite{Dokania2015}.
Irradiation at lower proton beam energy (12\,MeV) was carried out for independent verification of some of the channels by comparison of relative yields. Although the primary interest in the present work was to study the Aut rock samples, neutron activation for BWH rock was also carried out for comparison at one energy.
The neutron flux was estimated using the $^{56}$Fe(n,p)$^{56}$Mn reaction~\cite{124SnDokania2014}. For this purpose, the $^{nat}$Fe foils ($\sim$\,4.5 – 10\,mg/cm$^{2}$) were placed in front and back of the rock samples during each irradiation. 
The activity of $^{56}$Mn in the irradiated iron foils was extracted from the yield of 846.8\,keV gamma ray, which was measured with the D2 or TiLES. The yield for 846.8\,keV was corrected for losses due to coincident summing in the detector with 1810.7 keV ($I_\gamma$ = 26.9\%) or 2113.1\,keV ($I_\gamma$ = 14.2\%). The corrections due to coincident summing from other low intensity gamma rays  (branching ratio $< 1.5\%$) are found to be negligible. 
The energy averaged neutron flux $< \phi_n >$ is obtained using relation

\begin{equation}\label{eq:3}
< \phi_n > = \frac {\sum_{E_n} \sigma({E_n}) \phi_n(E_n) dE_n}{\sum_{E_n} \sigma({E_n}) dE_n}
\end{equation}
where $\sigma (E_n)$ is the neutron induced reaction cross section and $\phi_{n} (E_n)$ is the neutron flux  per unit energy at $E_{n}$. The numerator is derived from the measured $^{56}$Mn activity with appropriate corrections for the decay during irradiation and cooldown time.
The cross-sections for the $^{56}$Fe(n,p)$^{56}$Mn reaction were taken from the ENDF/B-VIII.0 library~\cite{ENDF}. The error in $<\phi_n>$ is largely limited by the statistical error in extracting the yield of 846.8\,keV gamma ray. It should be noted that the uncertainties in the neutron cross-sections have not been considered in the final errors. Table~\ref{tbl2} gives the details of the the rock samples and irradiation (average proton beam current, irradiation time $t_{irr}$, etc.)

The irradiated rock samples and iron foils were removed from the irradiation setup after a sufficient cool-down time $t_{c}$ ($\geq$\,1-2\,hr). Hence, some of the short-lived activities could not be measured. The offline counting of the irradiated rock samples was carried out in a close geometry in both setups and reaction products were identified by their characteristic gamma rays. In case of low energy irradiation, where both BWH and Aut rock samples were irradiated simultaneously for optimal use of the beam time, priority for counting in the off-line setups was given to the irradiated Aut samples. 
Spectra of rock samples  were recorded at different cool-down times ($t_c$) to track activities with varying half-lives, namely, $\sim$\,mins to few $\sim$\,days.
The D1-D2 setup was mainly used for detection of coincident gamma rays for confirmation of the nuclide identification. 

\begin{table}[width=1.0\linewidth,cols=4,pos=h]
\caption{Details of the irradiation of rock samples together with maximum neutron energy ($E_n$) and estimated energy averaged neutron flux.}\label{tbl2}
\begin{tabular*}{\tblwidth}{@{}CCCCCCC@{}}
\toprule
Sample   & Mass      & $E_{p}$  & $<I>$      &$t_{irr}$  & $E_{n}$      & $< \phi_{n} >$                          \\ 
         & (g)       & (MeV)    & (nA)       & (h)              & (MeV)        & ($10^6$~n\,cm$^{-2}$ s$^{-1}$\,MeV$^{-1}$)        \\ 
\midrule
AUT1A    & 10.6      & 22       & 136        & 15.5              & 19.9         & 1.9 (2)                             \\
\midrule
AUT4E    & 6.7  &\multicolumn{1}{c}{\multirow{2}{*}{12}} &\multicolumn{1}{c}{\multirow{2}{*}{148}}  &\multicolumn{1}{c}{\multirow{2}{*}{16.0}} &\multicolumn{1}{c}{\multirow{2}{*}{9.9}} & \multicolumn{1}{c}{\multirow{2}{*}{0.51 (4)}}\\ 
BWH2B & 5.5  &  &  &  &   &\\
\bottomrule
\end{tabular*}
\end{table}

\section {Analysis and Results}
\subsection{Radiopurity measurements}\label{Radiopurity measurements} 
\begin{figure}[h!]
\centering
\includegraphics[width=12.0cm,height=10.0cm]{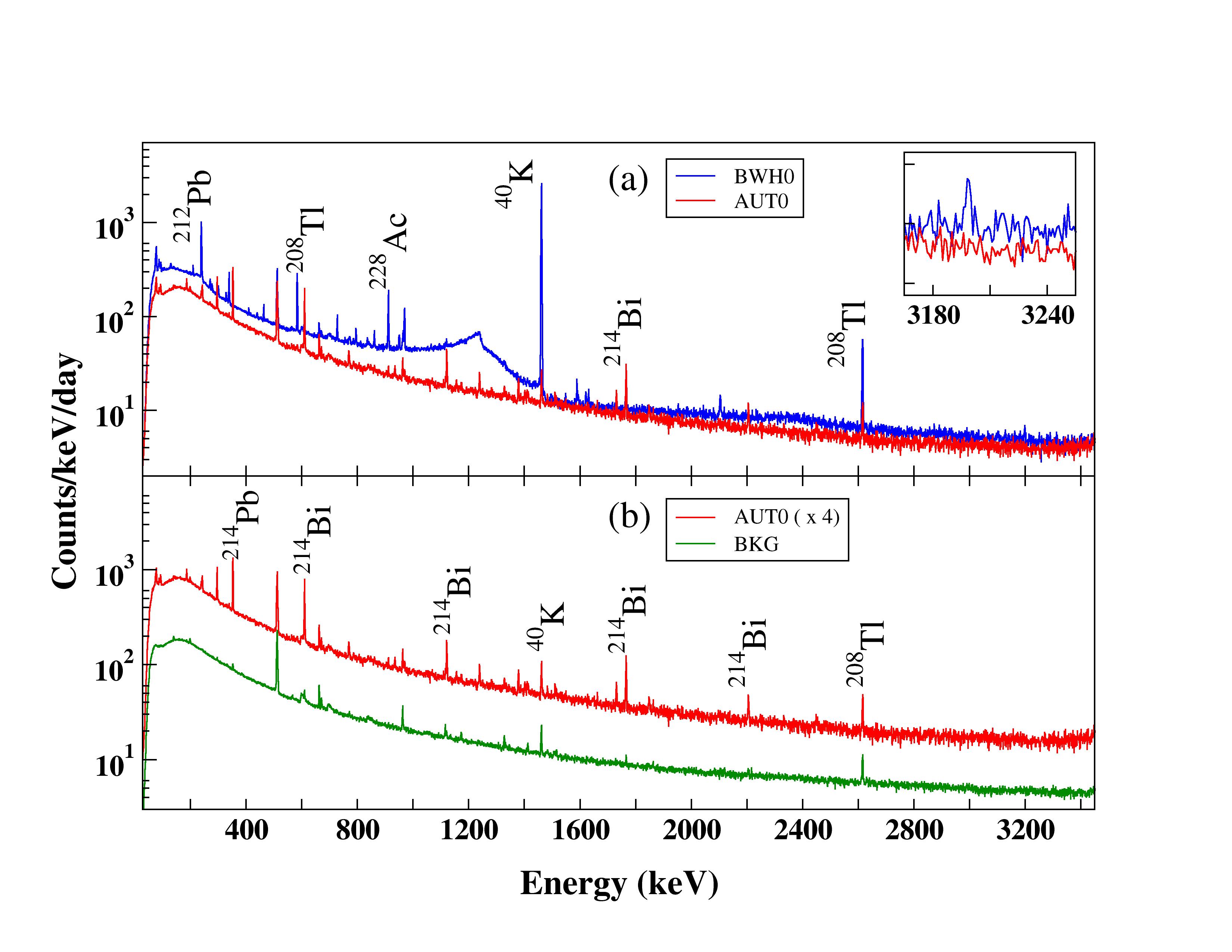}
\caption{Gamma ray spectra measured in TiLES for a) AUT0 (red line) and BWH0 (blue line),  b) ambient background (green line) and AUT0, scaled by an arbitrary factor of 4 for better visualisation. The inset in top panel shows the presence of the sum energy peak at 3197.7\,keV in the BWH rock.
}\label{fig2}
\end{figure}

A comparison of  gamma ray spectra of the AUT0 and BWH0 rock samples in TiLES is shown in Figure~\ref{fig2} along with the ambient background. Both the rock spectra are clearly dominated by gamma lines from the $^{238}$U and $^{232}$Th decay chains, as compared to the ambient background.  The high energy gamma rays from  $^{208}$Tl (end product in $^{232}$Th decay chain) - 2614.5\,keV ($I_{\gamma}$\,=\,99.75\,\%) are of particular concern. A  sum energy peak at  3197.7\,keV, arising from the coincident summing of 2614.5 and 583.2 keV is clearly visible in the BWH0 spectrum, but not observed in the AUT0 spectrum. The BWH0 also shows a strong peak at 1460.8\,keV, indicating large amount of $^{40}$K in the rock. This also results in significantly enhanced background at lower energies, $E$\,<\,1500\,keV, for the BWH0 as compared to the AUT0. 

Although many gamma rays from $^{232}$Th and $^{238}$U decay chains are visible, not all could be considered for trace impurity analysis. Some of the gamma rays are mixed from different radionuclides - for example, the observed gamma line at 242.6\,keV has contribution from $^{214}$Pb in $^{238}$U decay chain (242.0\,keV) and $^{224}$Ra in $^{232}$Th decay chain (241.9\,keV), similarly for 351.9\,keV ($^{214}$Pb, $^{211}$Bi), 185.9\,keV ($^{226}$Ra, $^{235}$U) etc. Hence, only  those gamma rays  which could be unambiguously assigned to a particular nuclide were considered in further analysis. Further, a subset was selected from coincident summing corrections as explained earlier.
To extract the peak area, the photopeaks of interest were fitted with a gaussian + background (second order polynomial). The specific activity, that is, activity per unit mass $A_E({\gamma})$ corresponding to a given transition of the radionuclide was determined using

\begin {equation}\label{eq:4} 
A_E ({\gamma}) = \frac{N_{obs}}{I_{\gamma}^{net}.\varepsilon_{\gamma}.m.t}
\end {equation}
where $N_{obs}$ is the net observed counts in the photopeak after correcting for the ambient background, $I_{\gamma}^{net}$ is the branching ratio of the gamma ray after summing correction, $\varepsilon_{\gamma}$ is the photopeak detection efficiency computed using GEANT4 simulation~\cite{Dokania2014}, \textit{m} is the mass of the sample and \textit{t} is the counting time. The $I_{\gamma}$ was corrected for coincident summing estimate of 2 gamma cascades, as explained earlier. However, the coincident summing is neglected if summing correction is significantly less than the statistical error.
In the cases where no measurable activity could be observed above the ambient background (at the present sensitivity of the setup), the upper limit on the specific activity ($L_{A}$) was estimated from the minimum detectable counts ($L_{D}$) using Currie's method~\cite{curie1968} as

\begin {equation}\label{eq:5} 
L_{A} = \frac{L_{D}}{I_{\gamma}.\varepsilon_{\gamma}.m.t}  
\end {equation}
where $L_{D}$ = 4.65\,$\sigma_{B}$ + 2.7 and $\sigma_{B}$ is the standard deviation in the background counts. The extracted specific activities for the AUT0 and BWH0 samples are listed in Table~\ref{tbl3}. Due to its small isotopic abundance, the concentration of $^{235}$U could not be determined in the present work.

\begin{table}[width=1.0\linewidth,cols=4,pos=h]
\centering
\caption{Observed specific activities for different radionuclides in the AUT0 and BWH0 rocks.}\label{tbl3}
\begin{tabular*}{\tblwidth}{@{}CCCCCCC@{}}
\toprule
Parent Radionuclide          & Daughter Radionuclide      & $E_{\gamma}$   &\multicolumn{2}{C}{AUT}     &\multicolumn{2}{C}{BWH}     \\
\midrule
                &             &                & $A_E$           &  $<A_E >$          &    $A_E$    & $<A_E>$                 \\
                 &              &(keV)           & (mBq/g)              & (mBq/g)             & (mBq/g)            & (mBq/g)                 \\
\midrule
$^{40}$K        &               &1460.8          &$<$\,1                &  $<$\,1           & 1064 (68)               & 1064 (68)                        \\
\midrule                                  
$^{232}$Th      &$^{212}$Pb     &238.6           & 0.50 (7)                 &   {\multirow{3}{*}{0.50 (6)}}               &    15 (1)                 &      {\multirow{3}{*}{14.7 (6)}}                 \\
                &$^{228}$Ac     &911.2           & 0.5 (2)              &               &     15 (1)                 &                      \\
                &               &969.0           & ---              &                &     14 (1)                &                      \\ 
\midrule                         
$^{238}$U       &$^{214}$Pb     &295.3           & 8.2 (5)             &  {\multirow{8}{*}{8.2 (3)}}              &      1.2 (2)               &  {\multirow{8}{*}{ 1.2 (1)}}                   \\
                &$^{214}$Bi     &609.3          & 7.7 (4)               &                 &     1.5 (2)                &                     \\
                   &    & 1120.3          & 8.1 (5)               &                &        ---             &                     \\             
                &               &1377.7          &12 (2)              &                    &    ---                 &                     \\
                &               &1729.6          &13 (2)              &                      &  ---                   &                    \\
               &                &1764.5          &7.8 (7)             &                   &     0.9(3)                &                     \\
                &               &1847.5          &14 (3)              &                      &     ---                &                    \\
                &               &2204.1          &10 (1)              &                     &      ---               &                    \\
\bottomrule
\end{tabular*}
\end{table} 
The mean specific activity (${<}\,A_E\,{>}$) is obtained by weighted average over measured specific activities of different gamma rays/daughter nuclides in the given decay chain.
In order to extract the concentrations of the parent radionuclides ($^{232}$Th and $^{238}$U), the data were analyzed under the assumption of secular equilibrium within samples. The atomic fraction $F_{E}$ of the trace radioimpurity in the rock sample was computed using 

\begin {equation}\label{eq:6} 
F_{E}\,(ppb) = \frac{A_{E}.M}{\lambda.N_{A}} . 10^{6} 
\end {equation}
where $M$ is the molar mass (in g/mole), $\lambda$ is the decay constant (in s$^{-1}$) and $N_{A}$ is Avogadro’s number.

For the BWH0 sample, the molar mass was determined from the rock composition obtained from the Secondary Ion Mass Spectroscopy (SIMS) results reported in Ref.~\cite{Dokania2015}. 
Further, in the case of the AUT0 sample, since the rock composition was not available, the molar mass was assumed to be identical to that of the BWH0. This is a reasonable assumption as the molar mass is not expected to vary significantly for different types of rocks. The concentrations of the trace radioimpurities for both the rock samples are listed in Table~\ref{tbl4}. 

\begin{table}[width=1.0\linewidth,cols=4,pos=h]
\centering
\caption{Trace radioimpurity concentrations in AUT0 and BWH0 rocks. }\label{tbl4}
\begin{tabular*}{\tblwidth}{@{}CLL@{}}
\toprule
Sample          &Parent Radionuclide         &  Concentration   \\
                &                     & (ppb)            \\
\midrule

AUT0            &$^{40}$K               & <2         \\ 
                &$^{232}$Th             & 12 (1)        \\
                &$^{238}$U              & 60 (2)        \\
\midrule

BWH0            &$^{40}$K               &2179 (139)    \\ 
                &$^{232}$Th             &338 (14)     \\
                &$^{238}$U              &9 (1)        \\
\bottomrule
\end{tabular*}
\end{table}

From the comparison of trace element concentrations in AUT0 and BWH0 in Table~\ref{tbl4}, it is evident that the $^{40}$K content in the AUT0 is significantly lower by a factor of $\sim$\,1000.  
The content of $^{232}$Th is also lower in the AUT0 by a factor of $\sim$\,28, whilst that for the $^{238}$U is higher by a factor of $\sim$\,7.
The concentrations of $^{40}$K and $^{232}$Th in the BWH0 rock obtained in the present study are similar to earlier reported values~\cite{Dokania2015}, namely, 2520\,ppb (by SIMS) and 224\,ppb (by Inductively Coupled Plasma Mass Spectroscopy i.e. ICPMS), respectively. However, for $^{238}$U the present value is significantly lower than that reported with ICPMS, namely 60\,ppb~\cite{Dokania2015}. It should be pointed out that radiopurity measurements reflect average over a larger finite sample size as compared to the ICPMS. Hence, the observed difference may arise due to the non-homogeneous distribution of radionuclides within the rock. The variation in trace impurity content was also probed by measurements on different BWH/Aut samples from the same location.  
While no differences were observed in the spectra at the measured level of sensitivity for  three different Aut rock samples ($t\sim$\,2 weeks), two BWH rock samples showed  $\sim33\,\%$ and $\sim17\,\%$ variation for $^{232}$Th and $^{40}$K, respectively.
 
As a result of lower levels of $^{40}$K and $^{232}$Th in AUT0, the yield of the high energy gamma rays of 1460.8\,keV and 2614.5\,keV is significantly reduced as compared to the BWH0. The higher $^{232}$Th content in BWH0 is also reflected in the presence of 3197.7\,keV gamma ray (see inset of Figure~\ref{fig2}).  
The presence of high energy gamma rays and the corresponding Compton background leads to an overall enhancement in background at lower energies, i.e. in the region of interest relevant to low background experiments ($E$\,$\sim$\,2\,MeV). Thus, the smaller $^{232}$Th content in the Aut rock appears to be advantageous.

The observed specific activities in the AUT0 (see Table~\ref{tbl3}) can be compared with those at the well established underground facility LNGS, which also has dolomitic limestone rock as well as the worldwide average. The measured average specific activities in LNGS~\cite{malczewski2013LNGS} are 26 (2)\,mBq/g for $^{40}$K, 1.5 (1.0)\,mBq/g for $^{232}$Th and 1.8 (1.0)\,mBq/g for $^{238}$U. Thus, the Aut rock appears to have lower levels of both $^{40}$K and $^{232}$Th, while $^{238}$U content is somewhat higher. It is important to note that the trace radioactive element content in Aut rock is significantly lower than the worldwide average, namely, 400, 30 and 35\,mBq/g for $^{40}$K, $^{232}$Th and $^{238}$U, respectively~\cite{UNSCEAR2000}. 
The neutron flux at the BWH site is estimated to be 2.76\,(47)\,$\times$\,10$^{-6}$\,n\,cm$^{-2}$\,s$^{-1}$ with 60\,ppb for $^{238}$U and 224\,ppb of $^{232}$Th~\cite{Dokania2015} trace impurities in the rock. The low energy neutron flux produced by spontaneous fission and ($\alpha$,n) reactions in the rocks is dominated by $^{238}$U. The measured concentration of $^{238}$U in the AUT0, 60\,(2)\,ppb, is similar to that used for the neutron flux estimation at the BWH site (namely, 60\,ppb). Consequently, the expected neutron flux at the Aut site will be similar to that for the BWH site, i.e. $\sim 3\times\,10^{-6}$ n\,cm$^{-2}$\,s$^{-1}$. It should be noted that this estimate is comparable to other underground laboratories, although concrete wall contributions will have to be taken into consideration at the actual site. Thus, Aut site is expected to have a overall lower gamma ray background and similar neutron background as compared to the BWH and subsequently appears to be a suitable site for laboratory from radioactive background considerations.

\subsection{Neutron activation measurements}\label{Neutron activation measurements}

\begin{figure}[h!]
\centering
\includegraphics[width=12.0cm,height=10.0cm]{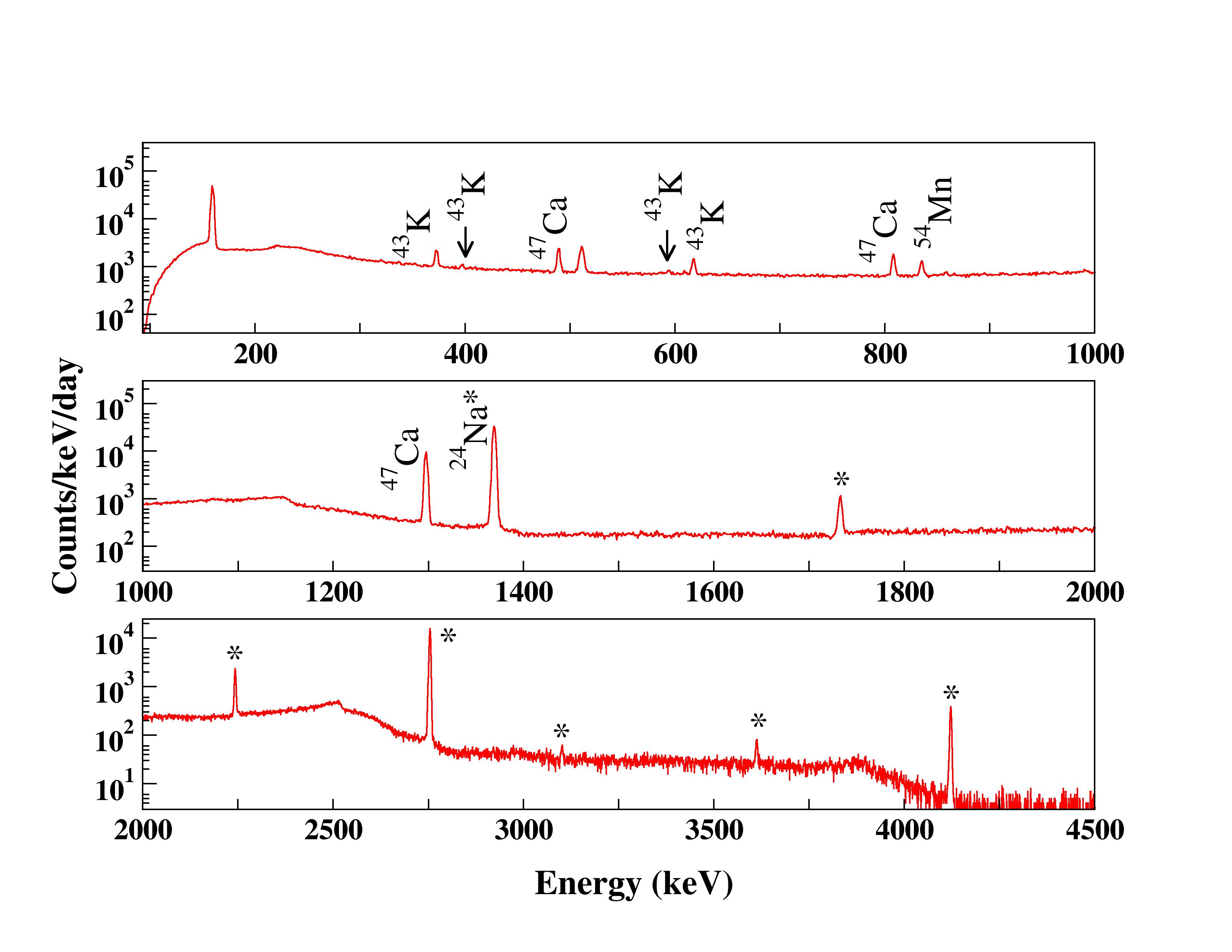}
\caption[irradiated spectra]{Gamma ray spectrum  of the irradiated AUT1A rock ($E_p$\,=\,22\,MeV) after $t_{c}$\,=\,5\,d. Various $^{24}$Na gamma rays and associated single/double escape peaks are indicated (*).}
\label{fig3}
\end{figure}

\begin{figure}[h!]
\centering
\includegraphics[width=12.0cm,height=10.0cm]{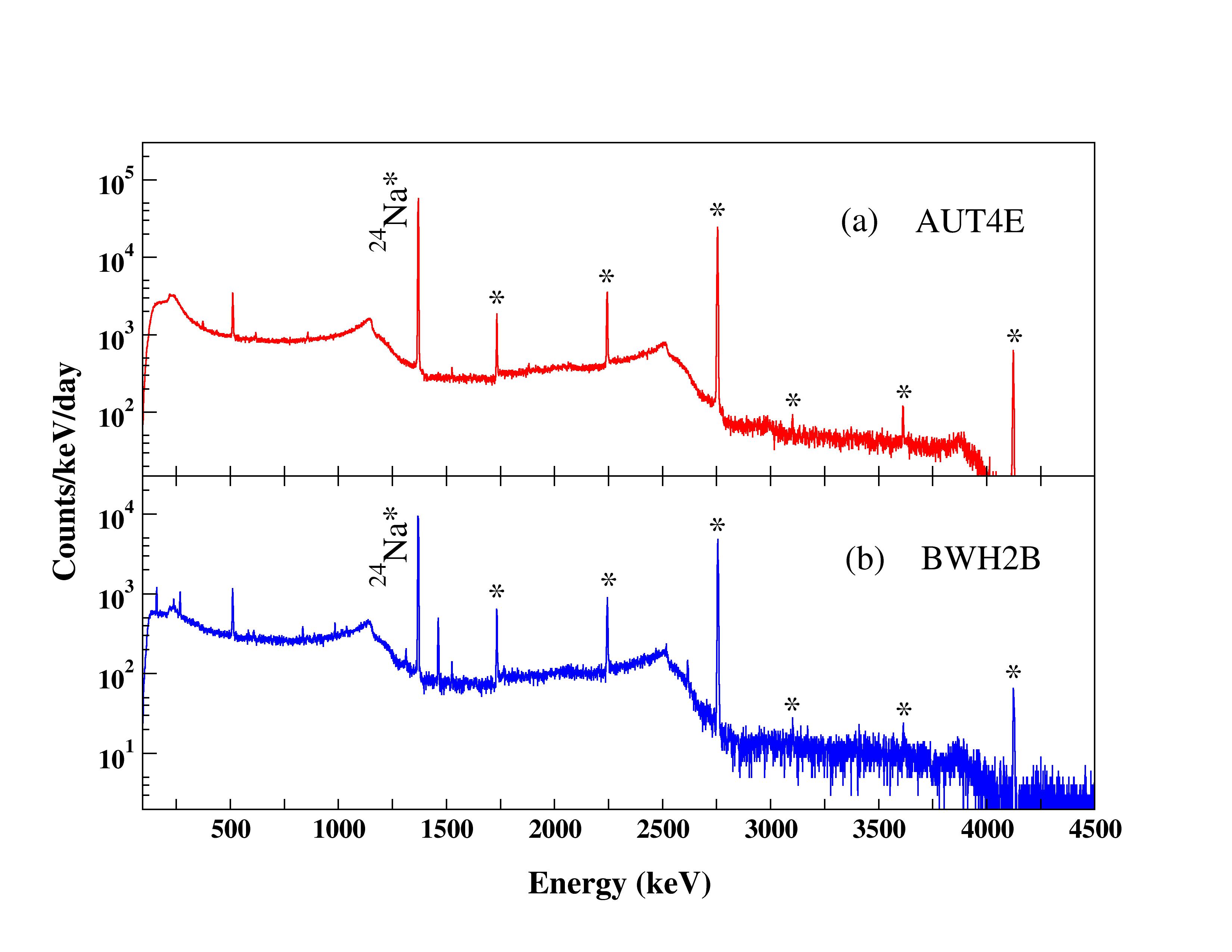}
\caption[irradiated spectra]{Gamma ray spectra of the irradiated ($E_p$\,=\,12\,MeV) rock samples after $t_{c}$\,=\,1.7\,d, (a) AUT4E in TiLES and (b) BWH2B in D1. The * mark has same meaning as in Figure~\ref{fig3}}.
\label{fig4}
\end{figure}

Figure~\ref{fig3} and \ref{fig4} show gamma ray spectra of the irradiated rocks at $E_p$\,=\,22 and 12\,MeV, respectively. Various observed reaction channels such as (n,$\gamma$), (n,p), (n,$\alpha$), and (n,2n) are listed in Table~\ref{tbl5} together with respective $T_{1/2}$ and prominent gamma rays. The threshold neutron energy $E_{th}$ corresponding to $\sim\,1\,\mu$b cross-section is also listed in the table. 

\begin{table}[width=1.0\linewidth,cols=4,pos=h]
\centering
\caption{Observed neutron-induced reaction products in Aut and BWH rocks together with the threshold neutron energy $E_{th}$ corresponding to $\sim\,1\,\mu$b cross section. The half-life ($T_{1/2}$) and prominent gamma rays ($E_{\gamma}$) are also listed.}\label{tbl5}
\begin{tabular*}{\tblwidth}{@{}LLLLL@{}}
\toprule
Reaction                        & $E_{th}$     & $T_{1/2}$       & $T_{1/2}$           & $E_{\gamma}$                                \\
channel                         & (MeV)       & (reference)     & (measured)          & (keV)                                          \\
\midrule 
$^{48}$Ti(n,p)$^{48}$Sc         & 5           & 1.82\,d         & ---                 &983.5, 1037.5, 1312.1                       \\
$^{48}$Ca(n,2n)$^{47}$Ca        & 10          &4.54\,d          &4.47 (6)\,d          &489.2, 807.9, 1297.1                               \\    
$^{24}$Mg(n,p)$^{24}$Na, $^{27}$Al(n,$\alpha$)$^{24}$Na         & 5, 4.64           &15\,h            &15.04 (5)\,h         &1368.6, 2754.0  \\ 
$^{43}$Ca(n,p)$^{43}$K          & 1.8         &22.3\,h          &22 (1)\,h            &373.8, 617.5                                     \\ 
$^{46}$Ti(n,p)$^{46}$Sc         & 3          &83.79\,d         & ---                 &889.3, 1120.5                                      \\
$^{23}$Na(n,2n)$^{22}$Na        & 13          &2.60\,y          & ---                 &511.0, 1274.5                                           \\      
$^{39}$K(n,2n)$^{38}$K          & 14.5        &7.64\,min        & ---                 &2167.5                                           \\      
$^{41}$K(n,p)$^{41}$Ar          & 3           &1.82\,h          & ---                 &1293.6                                              \\      
$^{54}$Fe(n,p)$^{54}$Mn         & 0.72        &312\,d           &279 (68)\,d          &835.0                                             \\ 
$^{56}$Fe(n,p)$^{56}$Mn         & 4           &2.58\,h          & ---                 &846.8                                               \\ 
$^{43}$Ca(n,n'p)$^{42}$K, $^{41}$K(n,$\gamma$)$^{42}$K      & 12, 1          &12.36\,h         &12.8 (1)\,h          &1524.7                                             \\
\bottomrule
\end{tabular*}
\end{table}

\begin{figure}[h!]
\centering
\includegraphics[width=8.1cm,height=6.4cm]{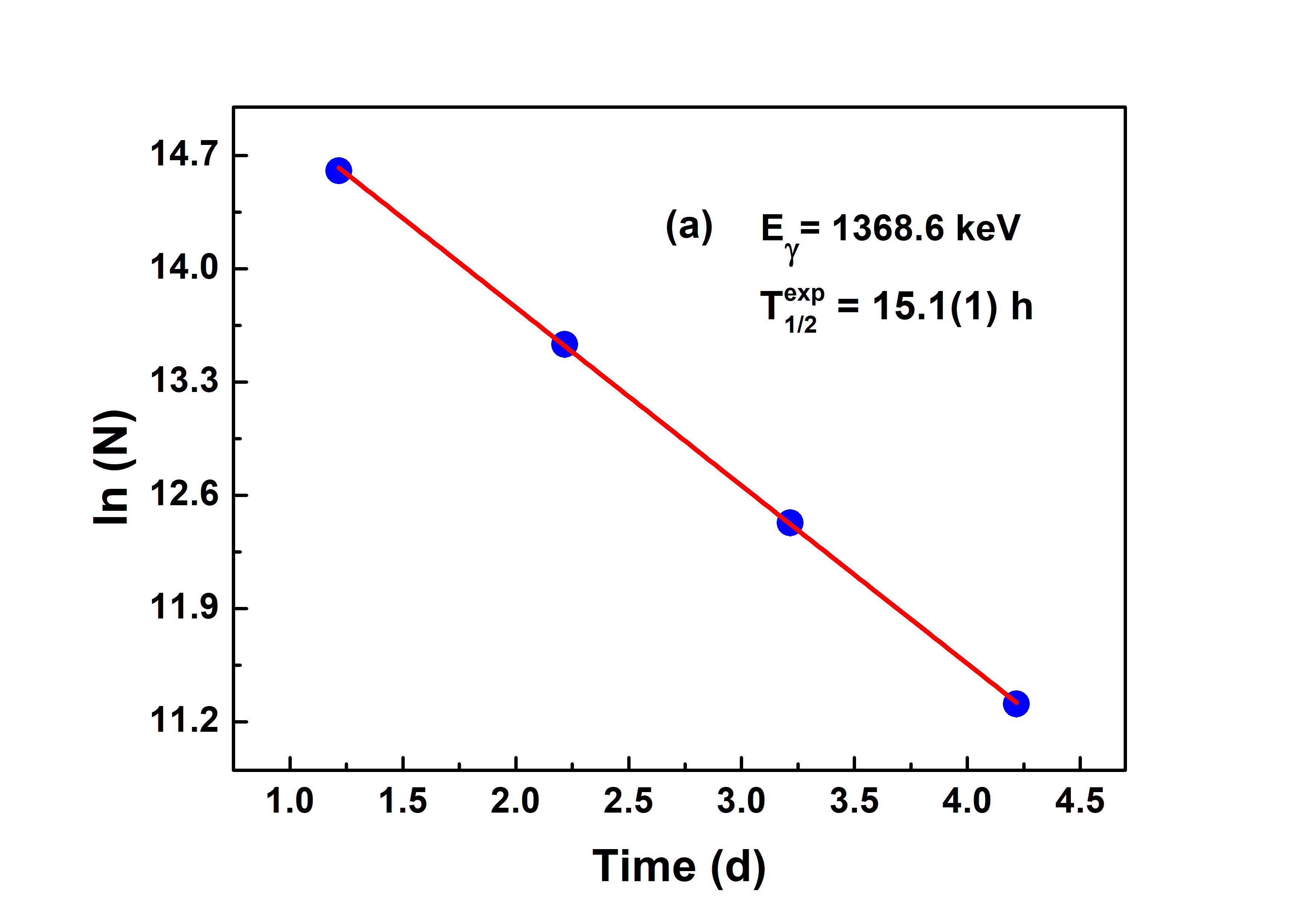}
\includegraphics[width=8.1cm,height=6.4cm]{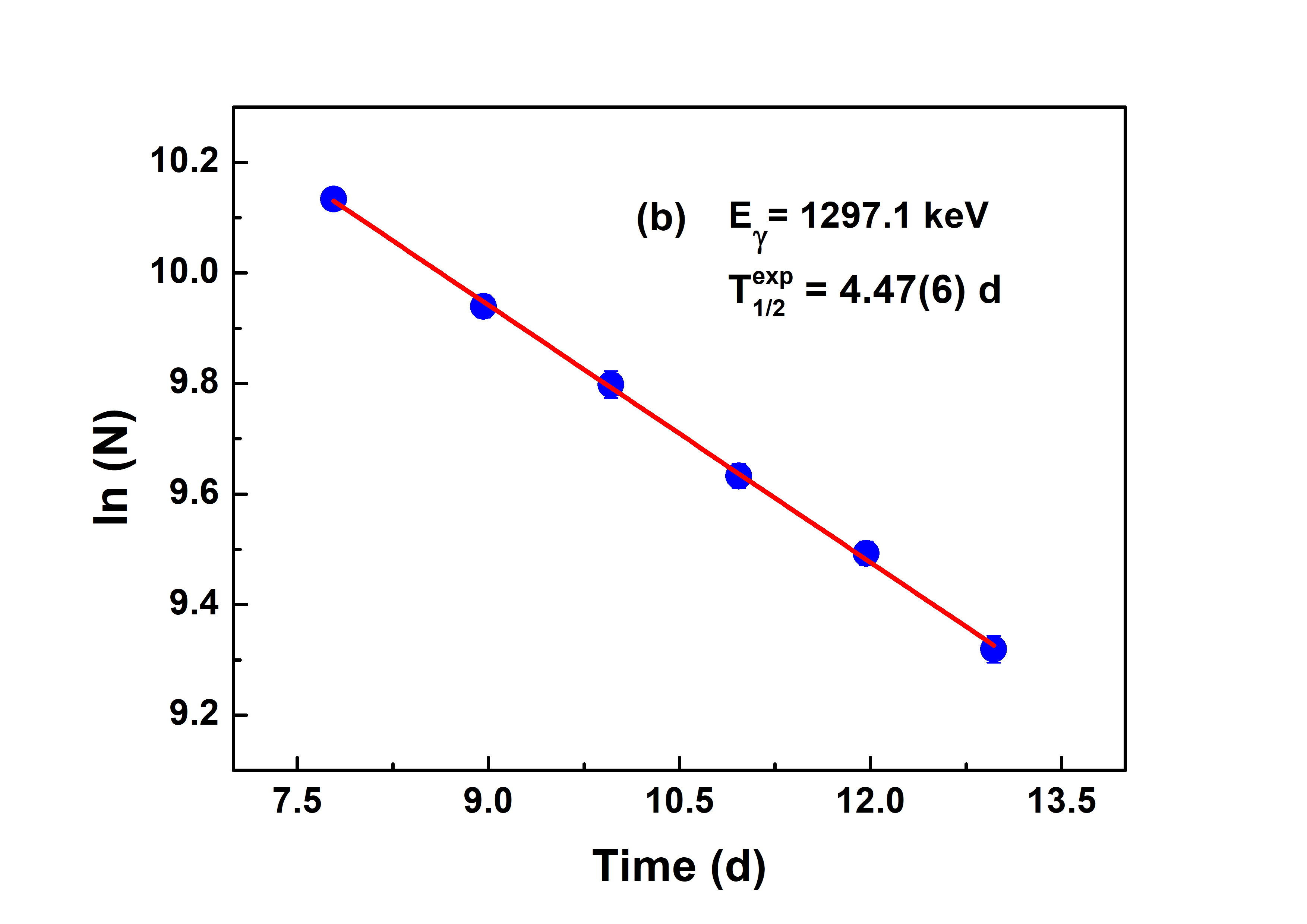}
\caption[irradiated spectra]{Decay curves  (a) $E_{\gamma}$\,=\,1386.6\,keV of $^{24}$Na and (b) $E_{\gamma}$\,=\,1297.1\,keV of $^{47}$Ca, where N are photopeak counts corresponding to  integration time of 3\,h and 24\,h for (a) and (b), respectively. Errors are smaller than the symbol size.}
\label{fig5}
\end{figure}

In the AUT1A spectrum, the dominant contribution at small $t_c$ (i.e. $\le2-3$ days) comes from the $^{24}$Na activity ($T_{1/2}$\,=\,15\,h). The high energy gamma rays, namely, 1368.6\,keV and 2754.0\,keV, and the respective single/double escape peaks are clearly visible in the figure. It is important to note that the associated Compton background also leads to the  enhancement of the low energy background. The $^{24}$Na can be formed either via $^{23}$Na(n,$\gamma$) or $^{24}$Mg(n,p) or $^{27}$Al(n,$\alpha$).  
With fast neutrons, $^{23}$Na(n,$\gamma$) is less probable as compared to $^{24}$Mg(n,p) or $^{27}$Al(n,$\alpha$). However, the relative contributions of these two channels will depend on the actual rock composition, namely, Al/Mg content. It is seen that the BWH rock contains more aluminum than magnesium (from the SIMS data). The Aut rock also shows the presence of  $^{42,43}$K and $^{47}$Ca  originating from the Calcium isotopes, which is present in dolomite rock. These reaction products are relatively short-lived ($T_{1/2}\le$ few days) and give rise to low energy gamma rays, $E_{\gamma}$\,$<$\,1300\,keV, during the decay. 
The long-lived activities $^{22}$Na ($T_{1/2}=2.6$\,y) and $^{54}$Mn ($T_{1/2}=0.855$\,y) were measured after sufficiently long cool-down time, which ensured that dominant short-lived products have diminished. Further, identification of long-lived activities $^{22}$Na and $^{46}$Sc were confirmed by the detection of coincident gamma rays in the decay cascade in D1-D2 setup. The measured half-life of the reaction products were found to be in good agreement with  the reference values~\cite{nndc}. Figure~\ref{fig5} shows typical decay curves for a couple of reaction products in the AUT1A sample ($E_p$\,=\,22\,MeV).  

From the measured yield of the characteristic gamma ray during time $t_{1}$ to $t_{2}$, the saturated activity $A_\infty$~\cite{knoll} (for $t_{irr} \rightarrow \infty$) can be obtained as 

\begin{equation}\label{eq:7}
A_\infty = \frac{\lambda N_{obs}}{\epsilon_\gamma I_\gamma^{net} (1-e^{-\lambda t_{irr}})e^{\lambda t_{irr}}(e^{-\lambda.t_{1}}-e^{-\lambda.t_{2}})}
\end{equation}
The saturated activity $A_\infty$ per unit mass has been estimated for the prominent reaction products in the Aut rock and are presented in Table~\ref{tbl6}.
The coincident summing corrections have  been taken into consideration as samples were counted in the close geometry (see eqn.~\ref{eq:2}) and were found to be around  10 - 16\,\% for the nuclides $^{22}$Na, $^{24}$Na and $^{43}$K.
The nuclides $^{22}$Na and $^{47}$Ca are absent in the AUT4E sample, which is irradiated at lower energy ($E_p$\,=\,12\,MeV). This is expected as both these nuclides are produced via (n, 2n) reaction which have higher threshold energies and hence require $E_n>$\,10\,MeV. The differences in the $A_\infty$ per unit mass at $E_p$\,=\,12 and 22\,MeV correspond to contribution from high energy neutrons (i.e. $E_n\sim$\,10-20\,MeV) and can also be estimated from Table~\ref{tbl6}. It is evident that production of both $^{24}$Na and $^{42,43}$K is dominated by the high energy neutrons.

\begin{table}[width=1.0\linewidth,cols=4,pos=h]
\centering
\caption{Saturated activity $A_\infty$ per unit mass for the prominent reaction products. The $E_{coin}$, emitted in coincidence with $E_{\gamma}$, which has been considered for the summing corrections are also listed.}\label{tbl6}
\begin{tabular*}{\tblwidth}{@{}LCCCC@{}}
\toprule
Nuclide          & $E_{\gamma}$  & $E_{coin}$  & $A_\infty$                        & $A_\infty$                      \\
                 & & & ($E_p$\,=\,12\,MeV)          &   ($E_p$\,=\,22\,MeV)    \\
                 & (keV) & (keV) & (Bq/g)                    & (Bq/g)                \\
\midrule
$^{22}$Na       & 1274.5 & 511.0 &  ---  &   9 (2)  \\ 
\midrule
{\multirow{2}{*}{$^{24}$Na}} & 1368.6 & 2754.0 & {\multirow{2}{*}{336 (14)}}  & {\multirow{2}{*}{6536 (299)}} \\ 
& 2754.0 & 1368.6 & & \\
 \midrule
$^{42}$K        & 1524.6 & --- & 4.6 (6)  & 62 (13) \\ 
\midrule
$^{43}$K        & 617.5 & 372.8 & 0.52 (8)  &  18 (1)  \\ 
\midrule
$^{47}$Ca       & 1297.1 & --- & ---  &  54 (3)  \\ 
\bottomrule
\end{tabular*}
\end{table}

It should be mentioned that  it is not possible to extract the concentration of the parent nuclides since the precise shape of the neutron energy distribution is not available in the present experiment. Moreover, in fast neutron induced reactions, multiple reaction pathways can lead to the production of a particular isotope (see Table~\ref{tbl5}). Hence, a thermal neutron activation study would be desirable for quantitative estimation of the trace elements and/or rock composition.

\begin{figure}[h!]
\centering
\includegraphics[width=12.0cm,height=10.0cm]{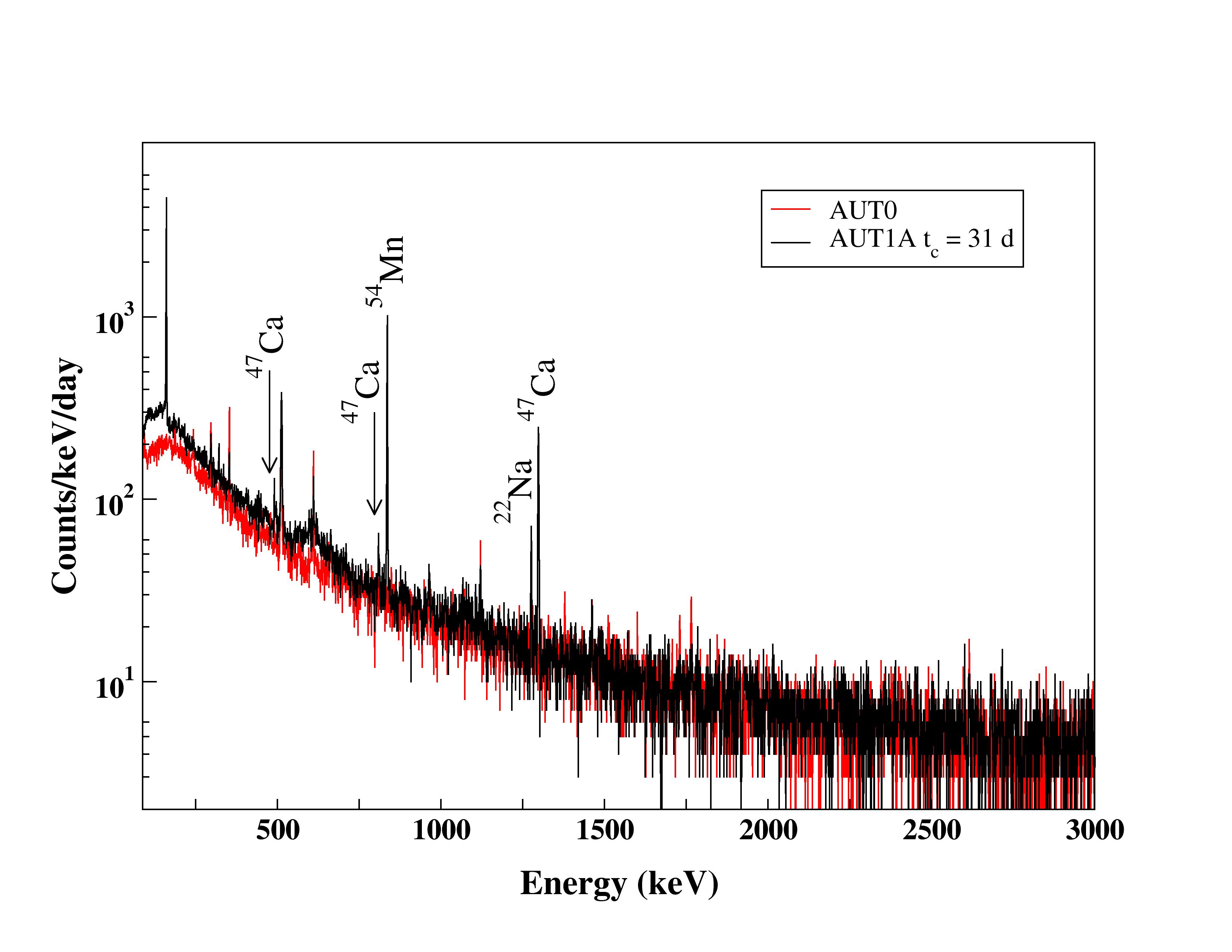}
\caption[irradiated spectra]{Comparison of gamma ray spectra of the irradiated ($E_p$\,=\,22\,MeV) AUT1A (black line) after $t_c$\,=\,31\,d and pristine AUT0 (red line) samples. Note that AUT1A (10.6\,g) has smaller mass than AUT0 (27.5\,g).}
\label{fig6}
\end{figure}

To understand the impact of long-lived activities, the spectrum of the irradiated Aut rock after $t_c\sim$\,30 days was compared with the spectrum prior to the irradiation. The effect of the remnant activities is reflected in the enhancement in the background at low energies as well as in the presence of  few characteristic gammas (100 - 1300\,keV) and can be seen in Figure~\ref{fig6}.

Overall, the BWH rock spectrum shows the presence of Potassium neutron induced reaction products, while that for Aut sample is dominated by Calcium products. The Aut rock appears to have mostly short-lived activities. However, long-lived products like $^{54}$Mn and $^{47}$Ca can result in the enhanced background at $E$\,$<$\,1000\,keV. The high energy gamma rays from $^{24}$Na need to be adequately shielded, as sum peak can  contribute to the background above 2\,MeV, which is region of interest for many NDBD studies.

\section{Summary}
\label{Summary and discussions}
The radiopurity studies of a rock sample from the potential underground laboratory site in the Aut tunnel have been carried out using the TiLES. The concentration of $^{40}$K in Aut rock is observed to be lower by a factor of $\sim$\,1000 as compared to the BWH rock sample. The measured specific activities of trace impurities $^{232}$Th and $^{238}$U in the Aut rock are 0.50\,(6) and 8.2\,(3)\,mBq/g, respectively. In comparison with BWH rock, the Aut rock appears to have lesser amount of $^{232}$Th and somewhat higher amount of $^{238}$U. It is important to note that Aut rock trace impurity concentrations are considerably lower than the respective worldwide average values. The low energy neutron flux arising due to spontaneous fission and ($\alpha$,n) reactions, dominated by $^{238}$U, is expected to be around 3\,$\times$\,10$^{-6}$ n\,cm$^{-2}$\,s$^{-1}$, which is similar to other underground laboratories.
The fast neutron activation studies of both Aut and BWH rock samples have indicated presence of long lived activities like $^{54}$Mn (0.855\,y) and $^{22}$Na (2.60\,y), but the resultant gamma ray energies are lower than 1500\,keV and no significant contributions at $E$\,$>$\,2\,MeV are observed. 
Overall, the ambient gamma ray background at Aut is expected to be lower than the BWH, while the low energy neutron background is expected to be similar. Hence, the Aut appears to be a suitable site for laboratory from radioactive background considerations.

\section*{Acknowledgement}
We are grateful to Project Director, NHAI, Pandoh-Aut-Takoli Tunnels section for helping us to collect the rock samples from the AUT tunnel (T02-02) at Mandi-Kullu National Highways. We thank Mr. K.V. Divekar for assistance during the measurements, the PLF staff for the smooth operation of the machine and the target lab staff for preparation of Fe foils. This work is supported by the Department of Atomic Energy, Government of India (GoI), under Project No. RTI4002. S. Thakur acknowledges the Ministry of Education, GoI, for Ph.D. research fellowship, and TIFR for supporting the visit related to this work. 

\bibliographystyle{elsarticle-num}

\bibliography{bibliography}

\end{document}